\documentclass[iop]{emulateapj-rtx4}

\usepackage{natbib}
\shorttitle{Collisionless Electron Heating in the Supernova Remnant Reverse Shock}
\shortauthors{Yamaguchi et al.}

\begin{document}

\title{New Evidence for Efficient Collisionless Heating of Electrons \\
	at the Reverse Shock of a Young Supernova Remnant}

\author{Hiroya Yamaguchi\altaffilmark{1,2,3}}
\email{hyamaguchi@cfa.harvard.edu}
\author{Kristoffer A.\ Eriksen\altaffilmark{4}}
\author{Carles Badenes\altaffilmark{5}}
\author{John P.\ Hughes\altaffilmark{6}}
\author{Nancy S.\ Brickhouse\altaffilmark{3}}
\author{Adam R.\ Foster\altaffilmark{3}}
\author{Daniel J.\ Patnaude\altaffilmark{3}}
\author{Robert Petre\altaffilmark{1}}
\author{Patrick O.\ Slane\altaffilmark{3}}
\author{Randall K.\ Smith\altaffilmark{3}}

\altaffiltext{1}{NASA Goddard Space Flight Center, Code 662, Greenbelt, MD 20771, USA}
\altaffiltext{2}{Department of Astronomy, University of Maryland, College Park, MD 20742, USA}
\altaffiltext{3}{Harvard-Smithsonian Center for Astrophysics, 60 Garden Street, 
	Cambridge, MA 02138, USA}
\altaffiltext{4}{Los Alamos National Laboratory, P.O.\ Box 1663, Los Alamos, 
	NM 87545, USA}
\altaffiltext{5}{Department of Physics and Astronomy and Pittsburgh Particle Physics, 
Astrophysics and Cosmology Center (PITT PACC), University of Pittsburgh, 3941 O'Hara St, 
Pittsburgh, PA 15260, USA}
\altaffiltext{6}{Department of Physics and Astronomy, Rutgers University, 136 Frelinghuysen Road, 
Piscataway, NJ 08854, USA}

\begin{abstract}

Although collisionless shocks are ubiquitous in astrophysics, 
certain key aspects of them are not well understood. 
In particular, the process known as collisionless electron heating, 
whereby electrons are rapidly energized at the shock front, 
is one of the main open issues in shock physics.
Here we present the first clear evidence for efficient collisionless electron heating 
at the reverse shock of Tycho's supernova remnant (SNR), revealed by Fe-K diagnostics 
using high-quality X-ray data obtained by the \textit{Suzaku} satellite. 
We detect K$\beta$ ($3p$$\rightarrow$$1s$) fluorescence emission from 
low-ionization Fe ejecta excited by energetic thermal electrons at the reverse shock front, 
which peaks at a smaller radius than Fe-K$\alpha$ ($2p$$\rightarrow$$1s$) 
emission dominated by a relatively highly-ionized component.
Comparison with our hydrodynamical simulations implies instantaneous 
electron heating to a temperature 1000 times higher than expected from 
Coulomb collisions alone. 
The unique environment of the reverse shock, which is propagating with 
a high Mach number into rarefied ejecta with a low magnetic field strength, 
puts strong constraints on the physical mechanism responsible for this heating, 
and favors a cross-shock potential created by charge deflection at the shock front. 
Our sensitive observation also reveals that the reverse shock radius of this SNR is 
about 10\% smaller than the previous measurement using the Fe-K$\alpha$ morphology 
from the {\it Chandra} observations. 
Since strong Fe-K$\beta$ fluorescence is expected only from low-ionization plasma 
where Fe ions still have many $3p$ electrons, this feature is key to diagnosing 
the plasma state and distribution of the immediate postshock ejecta in a young SNR.

\end{abstract}

\keywords{shock waves  --- plasmas --- atomic data --- hydrodynamics
--- ISM: individual objects (SN\,1572; Tycho's SNR) --- ISM: supernova remnants
--- X-rays: ISM}

\section{Introduction}

A supersonic flow colliding with another medium will form a shock wave, 
which can be described as a discontinuity in the physical conditions of the flow.  
Shock waves have been extensively observed on Earth, and also in 
a number of astrophysical settings with a wide range of spatial scales: 
from the solar wind \citep[e.g.,][]{Tidman71,Schwartz88} 
to afterglows of gamma-ray bursts (GRBs) \citep[e.g.,][]{Meszaros97,Sari98} 
and merging galaxy clusters \citep[e.g.,][]{Markevitch05,Markevitch07}. 
Unlike terrestrial shocks, the low densities found in some astrophysical environments 
imply that the shock transition occurs on scales much shorter than the typical 
particle mean free path for Coulomb collisions.  Therefore, collisionless processes 
involving collective interactions between particles and electromagnetic fields 
must be responsible for the shock formation \citep{McKee74}. 
Despite the ubiquity and importance of these collisionless shocks in astrophysics, 
the detailed physical processes that determine their fundamental properties are 
still poorly understood.

One particularly mysterious process, which might be closely related to 
the formation of the collisionless shock front itself, is the rapid heating 
of electrons at the shock transition.
A strong shock with velocity $v_s$ will result in a downstream temperature 
$T_i = 3\,m_i\, v_s^2/ 16\,k_{\rm B}$, where $m_i$ is the mass of particle $i$ 
and $k_{\rm B}$ is the Boltzmann constant. 
Since the timescale for collisional equilibration between different species is much 
longer than the time a particle spends in the shock transition zone \citep{Spitzer62}, 
the equation above can be applied independently to each species $i$, 
unless collisionless processes contribute to the temperature equilibration. 
It follows that the electron temperature ($T_e$) should be much lower than 
the temperature of heavier ions ($T_{ion}$) immediately behind the shock, 
and they will slowly equilibrate to a common temperature via Coulomb collisions 
further downstream. However, a number of theoretical investigations have 
suggested that rapid collisionless electron heating can occur at shock fronts 
\citep[e.g.,][]{McKee74,Cargill88,Laming00,Ghavamian07}.

Supernova remnants (SNRs) offer an ideal site to study this heating observationally, 
because they form long-lived, fast shocks in both the interstellar medium (ISM) and 
the supernova ejecta. 
To date, most work has concentrated on Balmer-dominated shocks 
associated with SNR blast waves expanding into the ISM 
\citep[e.g.,][]{Raymond83,Laming96,Ghavamian01,Ghavamian07,Rakowski03,Helder11}. 
Studies of electron heating in reverse shocks (RSs) which are propagating into 
the supernova ejecta are, on the other hand, very limited \citep{Hamilton97,France11}. 
In this paper, we focus on the RS of Tycho's SNR, the remnant of 
the Type Ia supernova observed in A.D.\,1572.

There are a number of important differences in environment between blast waves and RSs. 
Unlike the ISM, supernova ejecta generally have only a small fraction of neutral particles, 
because the strong UV/soft X-ray flux from the shocked material can easily 
photoionize the heavy elements of the unshocked ejecta \citep[e.g.,][]{Hamilton97}.
Furthermore, in a Type Ia SNR the ejecta consist only of heavy elements, 
with no contribution of hydrogen and helium. 
The magnetic field strength expected in the interior of the white dwarf before 
the explosion ($B \lesssim 10^{13}$\,G) \citep{Suh00} will be dramatically 
diluted by expansion. 
From flux conservation, a conservative upper-limit for the field strength 
present near the RS location in Tycho's SNR is estimated to be $\sim$$10^{-7}$\,G. 
This is sufficiently high that the ion gyroradius, 
$r_{\rm g} \sim 10^{13}\,(A/56)\,(B/10^{-7}\,{\rm [G]})^{-1}$\,cm 
(where $A$ is the mass number), is much smaller than the SNR radius of 
$\sim$$10^{19}$\,cm, ensuring the formation of a collisionless RS. 
Yet the inferred strength is at least an order of magnitude below the typical value in 
the ISM (a few times $10^{-6}\,{\rm G}$). The geometry of the interior magnetic field 
should be highly ordered as the field is stretched radially by the expanding ejecta.  
These distinct properties, strongly contrasting with those in the ISM, allow us to 
probe  the physics of collisionless electron heating under conditions far different 
from any earlier work.

Previous hydrodynamical calculations applied to Tycho's SNR required 
the presence of a modest amount of collisionless electron heating at the RS to 
explain the observed X-ray flux from the shocked ejecta \citep{Badenes05,Badenes06}. 
This is in contrast to earlier results on another Type Ia SNR, SN\,1006, where 
little evidence for collisionless electron heating was found \citep{Hamilton97}.
Since the RS in Tycho's SNR has begun to propagate into the Fe-dominated ejecta 
\citep[e.g.,][]{Hwang98,Warren05,Badenes06}, the emission lines from Fe directly probe 
the conditions (i.e., electron temperature) in the postshock region. 
In general, the initial shock-heated material in SNRs is at a very low charge state 
and only gradually becomes ionized by collisions with hot free electrons. 
The ejecta in young SNRs are, therefore, in a non-equilibrium ionization (NEI) 
condition \citep{Masai84}. Collisional interactions between hot electrons and 
low-ionized Fe in an NEI plasma produce inner K-shell ionization of the Fe ions, 
followed by K$\alpha$($2p$$\rightarrow$$1s$) or K$\beta$ ($3p$$\rightarrow$$1s$) 
fluorescence transitions \citep[e.g.,][]{Palmeri03,Mendoza04}. 
These emission lines are excellent diagnostics of the temperature and 
charge population in the shocked material.

Here we present new, strong evidence for collisionless electron heating at the RS 
of Tycho's SNR, revealed by Fe K-shell X-ray lines from sensitive observations 
with the X-ray Imaging Spectrometer (XIS) onboard the \textit{Suzaku} satellite. 
Our observations allow the first-ever detailed study of weak Fe-K$\beta$ emission 
alongside the stronger Fe-K$\alpha$ line.
In \S2, we analyze observational data based on the state-of-the-art atomic physics 
models we compute.
In \S3, we constrain the efficiency of collisionless electron heating 
by comparing our hydrodynamical calculations, 
and finally we discuss the plausible mechanism of this efficient heating process.

\section{Observational Results}

\begin{figure}[t]
  \begin{center}
	\vspace{2mm}
	\includegraphics[width=8.2cm]{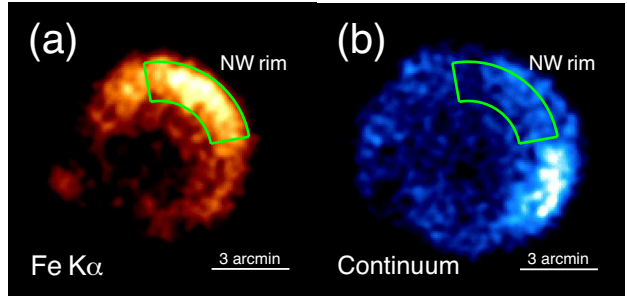}	
	\vspace{2mm}
\caption{Suzaku XIS images of Tycho's SNR in the (a) 6.43--6.53\,keV 
(Fe-K$\alpha$) band, and (b) 7.7--9.0\,keV band (continuum emission). 
North is up and east is to the left. 
The northwest (NW) region confined with the green lines is 
where we extract the spectrum shown in Figs.\,\ref{fig:spec1} and \ref{fig:spec2}. 
 \label{fig:image1}}
  \end{center}
\end{figure}

\begin{figure}[t]
  \begin{center}
	\includegraphics[width=8.2cm]{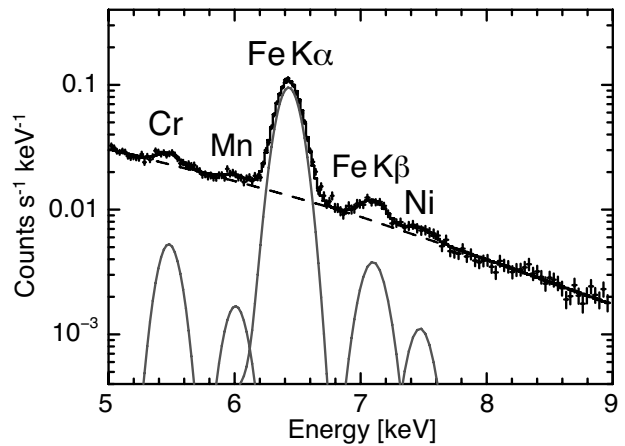}	
	\vspace{1mm}
\caption{
XIS spectrum of Tycho's SNR in the 5.0--9.0\,keV band, 
where the background data are subtracted. 
A phenomenological model using a power-law (dashed line for the continuum)
and five Gaussians (solid gray lines for the emission lines) yields the best-fit 
parameters given in Table\,\ref{tab:best}. 
  \label{fig:spec1}}
  \end{center}
\end{figure}

The observations with the {\it Suzaku} XIS were performed 
during August 2008 with a total effective exposure of 415\,ksec.
The primary data reduction is performed in accordance with 
the standard procedure recommended by the instrument team. 
We use only the data of front-illuminated CCDs (XIS0 and 3) and 
merge them to improve the photon statistics. 
Fig.\,\ref{fig:image1} shows XIS images in the energy bands of the Fe-K$\alpha$ line 
(a: 6.43--6.53\,keV) and continuum emission (b: 7.7--9.0\,keV). 
This continuum has been shown to be dominated by synchrotron radiation from 
relativistic electrons accelerated at the blast wave \citep{Hwang02,Cassam07}.
We extract an X-ray spectrum from the northwest (NW) rim indicated in Fig.\,1. 
Background data are taken from the nearby sky and subtracted from the source spectrum. 
Fig.\,\ref{fig:spec1} is the resultant spectrum in the 5.0--9.0\,keV band, where emission 
from Cr-K$\alpha$, Mn-K$\alpha$, Fe-K$\alpha$ and K$\beta$, and Ni-K$\alpha$ are 
clearly resolved. We note that this is the first detection of Ni 
from this remnant, although we do not focus on this element in this paper.

\subsection{Fe-K Emission Diagnostics}

We measure the centroid energies of the Fe-K$\alpha$ ($E_{\rm K\alpha}$) 
and K$\beta$ ($E_{\rm K\beta}$) line blends and their flux ratio ($\mathcal{R}$) 
using Gaussian functions, and obtain 
$E_{\rm K\alpha} = 6435 \pm 1$ $(\pm 6)$~eV, 
$E_{\rm K\beta} = 7104 \pm 10$ $(\pm 7)$~eV, 
and $\mathcal{R} =5.5_{-0.5}^{+0.6}$\%. The parenthetical values are 
the instrumental systematic uncertainty in the energy scale, 0.1\% of the mean 
energy, due to the incomplete gain calibration of the XIS \citep{Ozawa09}. 
The continuum is simply modeled by a power-law, giving the photon index of 
$\Gamma \sim 2.9$. 
The best-fit parameters for all the emission lines are given in Table\,\ref{tab:best}. 
During the analysis, an absorption column density of $N_{\rm H} = 7 \times 10^{21}$\,cm$^{-2}$ 
\citep{Cassam07} with the solar elemental composition \citep{Wilms00} is assumed, 
although the analyzed energy band is not affected by the foreground extinction.  
To evaluate the systematic uncertainty due to background subtraction, 
we perform fits using a different background dataset which includes only 
the instrumental component (non X-ray background). No significant 
change is found in the measurement of the line centroids and intensities.

\begin{table}[t]
\begin{center}
\caption{The best-fit spectral parameters for the NW rim.
  \label{tab:best}}
  \begin{tabular}{lcccc}
\hline \hline
Emission & Centroid & FWHM & Flux \\
~ & (eV) & (eV) & (10$^{-6}$\,ph\,cm$^{-2}$\,s$^{-1}$) \\
\hline
Cr K$\alpha$ & $5482_{-12}^{+10}$ & $141_{-45}^{+35}$ & $5.05_{-0.69}^{+0.72}$ \\
Mn K$\alpha$ & $6012_{-26}^{+25}$ & $141$ & $1.73_{-0.46}^{+0.47}$ \\
Fe K$\alpha$ & $6435 \pm 1$ & $138 \pm 2$ & $107 \pm 1$ \\
Fe K$\beta$  & $7104 \pm 10$ & $160 \pm 42$ & $5.62_{-0.56}^{+0.61}$ \\
Ni K$\alpha$ & $7478 \pm 32$ & $138$ & $1.82_{-0.43}^{+0.42}$ \\
\hline
\end{tabular}
\tablecomments{
The uncertainties are the statistical component in the 1$\sigma$ confidence range. 
The Gaussian widths (FWHM) of the Mn-K$\alpha$ Ni-K$\alpha$ lines 
were linked to those of Cr-K$\alpha$ and Fe-K$\alpha$, respectively. 
}
\end{center}
\end{table}

\begin{figure}[t]
  \begin{center}
	\includegraphics[width=8.2cm]{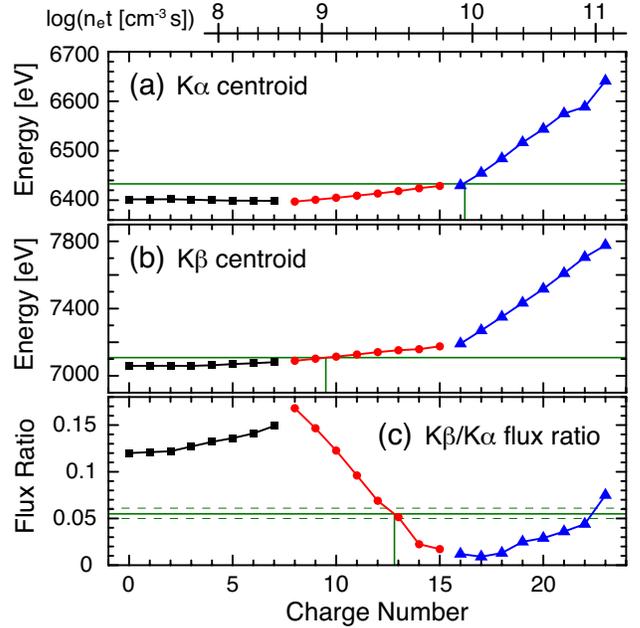}
		\vspace{1mm}
\caption{Expected centroid energies of the (a)\,Fe-K$\alpha$ 
and (b)\,Fe-K$\beta$ emission, and (c)\,the K$\beta$/K$\alpha$ flux ratio 
as a function of the charge number $z$ of Fe ions, with the corresponding ionization 
ages ($n_e t$) indicated at the top.
The best-fit values for Tycho's SNR are shown with the solid green lines. 
The dashed green lines in panel (c) indicate the $1\sigma$ lower- and upper-limits 
of the observed value. 
The black squares and red circles are the values calculated 
using the atomic data of \cite{Palmeri03} and \cite{Mendoza04}, 
respectively.  We also used the ``Flexible Atomic Code (FAC)'' \citep{Gu08} 
to calculate the rate coefficients of collisional ionization and excitation. 
The blue triangles are obtained by full calculations using the FAC 
(Eriksen et al., in preparation).
  \label{fig:diagnostics}}
  \end{center}
\end{figure}

The observed Fe-line parameters, $E_{\rm K\alpha}$, $E_{\rm K\beta}$, and $\mathcal{R}$, 
are compared in Fig.\,\ref{fig:diagnostics} to the theoretically-expected values for 
different charge numbers $z$ (where $z = 1$ indicates singly-ionized Fe) which are 
also tabulated in Table\,\ref{tab:values}. 
For $z = 0 - 7$ and $z = 8 - 16$, we use level energies, Einstein $A$-values, 
and fluorescence yields provided in the archival database of 
\cite{Palmeri03} and \cite{Mendoza04}, respectively. 
Since the transition probabilities of forbidden processes (e.g., $2s$$\rightarrow$$1s$) 
are negligible in multiple-electron ions \citep{Palmeri03,Mendoza04}, 
we take into account only $2p$$\rightarrow$$1s$ and $3p$$\rightarrow$$1s$ 
transitions as radiation channels for the K$\alpha$ and K$\beta$ emissions. 
We calculate rate coefficients for collisional ionization and excitation 
for each charge number using the ``Flexible Atomic Code (FAC)'' \citep{Gu08}. 
For $z \geq 16$, we perform full computational calculations with the FAC 
to obtain the theoretical values (Eriksen et al., in preparation). 
During the calculations, we assume an electron temperature of 5\,keV. 
The temperature-dependence is found to be significant only for $z \geq 20$, 
where the inner K-shell excitation rate becomes dominant over the K-shell 
ionization rate. 
Since the population of such highly charged Fe is not substantial in Tycho's SNR 
as is discussed below, the diagnostics we perform here are essentially 
independent of the electron temperature of the Fe ejecta.

\begin{table*}
\begin{center}
\caption{Theoretical values of the Fe-K$\alpha$ and K$\beta$ centroid energies 
and their intensity ratios for the different charge numbers $z$.
  \label{tab:values}}
  \begin{tabular}{ccccccccc}
\hline \hline
$z$ &$E_{{\rm K}\alpha}$ (eV) & $E_{{\rm K}\beta}$ (eV) & $\mathcal{R}$ & \ \ \ \ \ \ 
& $z$ & $E_{{\rm K}\alpha}$ (eV) & $E_{{\rm K}\beta}$ (eV) & $\mathcal{R}$ \\
\hline
0 & 6402 & 7059 & 0.120  && 12 & 6414 & 7141 & 0.069 \\ 
1 & 6402 & 7060 & 0.121  && 13 & 6419 & 7153 & 0.052 \\ 
2 & 6402 & 7060 & 0.122  && 14 & 6425 & 7159 & 0.022 \\
3 & 6401 & 7059 & 0.127  && 15 & 6428 & 7176 & 0.010 \\ 
4 & 6400 & 7063 & 0.132  && 16 & 6427 & 7192 & 0.012 \\ 
5 & 6399 & 7070 & 0.136  && 17 & 6455 & 7270 & 0.009 \\
6 & 6399 & 7075 & 0.141  && 18 & 6484 & 7351 & 0.013 \\ 
7 & 6399 & 7081 & 0.149 && 19 & 6517 & 7434 & 0.025 \\ 
8 & 6398 & 7090 & 0.168 && 20 & 6544 & 7517 & 0.029  \\ 
9 & 6401 & 7102 & 0.146 && 21 & 6575 & 7610 & 0.036  \\ 
10 & 6405 & 7115 & 0.122 && 22 & 6589 & 7705 & 0.044  \\ 
11 & 6410 & 7128 & 0.096 && 23 & 6641 & 7777 & 0.075  \\ 
\hline
\end{tabular}
\end{center}
\end{table*}

As found in Fig.\,\ref{fig:diagnostics}a, the observed $E_{\rm K\alpha}$ value 
corresponds to the charge states Fe$^{15+}$$\sim$Fe$^{17+}$
and an ionization age ($n_e t$) of $\sim 1\times 10^{10}$\,cm$^{-3}$\,s, 
where $n_e$ and $t$ are the electron density and the time elapsed since shock heating.  
This result is consistent with several previous measurements 
\citep[e.g.,][]{Hwang98,Hayato10}. It is frequently assumed that all the shocked ejecta  
responsible for the Fe K-shell emission have this ionization age. 
We find in Fig.\,\ref{fig:diagnostics}b, however, that the observed $E_{\rm K\beta}$ 
value corresponds to significantly 
lower charge states Fe$^{8+}$$\sim$Fe$^{10+}$, with an ionization age of 
$\sim 1\times 10^9$\,cm$^{-3}$\,s, about ten times lower than that indicated by 
the Fe-K$\alpha$ centroid. The K$\beta$ to K$\alpha$ flux ratio ($\mathcal{R}$) is 
also sensitive to the charge number, especially in the range $z = 8 - 14$ 
(Fig.\,\ref{fig:diagnostics}c). 
In this regime, the flux ratio experiences a drastic decrease because the Fe ions 
lose their $3p$-shell electrons (which are responsible for the K$\beta$ fluorescence) 
as $z$ increases.  The observed value is closest to the expected ratio for $z = 13$, 
intermediate between the results from the K$\alpha$ and K$\beta$ centroids.

\begin{figure}[t]
  \begin{center}
		\vspace{2mm}
	\includegraphics[width=8.2cm]{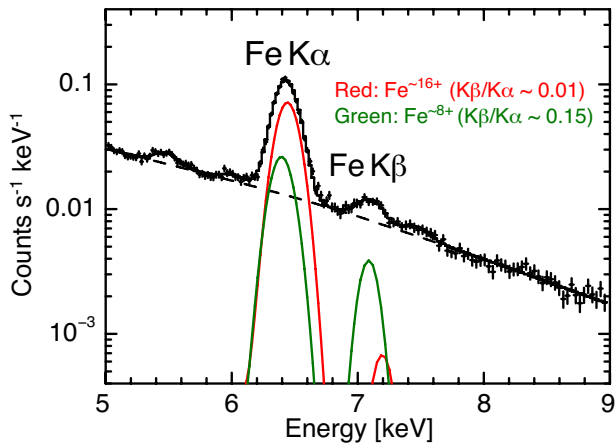}	
	\vspace{1mm}
\caption{Same spectral data as in Fig.\,\ref{fig:spec1}, but 
the `two-component' model is applied to the Fe-K emission (see text for details). 
Red and green represent the higher-ionization (around Fe$^{16+}$) and 
lower-ionization (around Fe$^{8+}$) components, respectively. 
The K$\beta$/K$\alpha$ flux ratio is expected to be lower in the former  
($\sim$0.01) than the latter ($\sim$0.15), because Fe$^{16+}$ ions 
have no bound electron in the $3p$ shell, which is responsible for the K$\beta$ emission. 
  \label{fig:spec2}}
  \end{center}
\end{figure}

The inconsistency among the three diagnostics indicates the presence of 
a range of plasma conditions, with the K$\alpha$ emission being dominated by 
more highly ionized and the K$\beta$ emission by less ionized Fe. 
We re-fit the NW spectrum applying a `two-component' model for 
the Fe emission. The red Gaussians in Fig.\,\ref{fig:spec2} correspond to the higher ionization 
component, where $E_{\rm K\beta}$ and $\mathcal{R}$ are fixed to be 7200\,eV 
and 1\% (the values theoretically expected for $z \sim 16$).
Only the K$\alpha$ centroid is allowed to vary, yielding 
$E_{\rm K\alpha} = 6447_{-3}^{+2}$~eV, which is in between the values 
for Fe$^{16+}$ and Fe$^{17+}$. 
The contribution of the lower ionization component is indicated by 
the green Gaussians in Fig.\,\ref{fig:spec2}, where we fix $E_{\rm K\alpha}$ and 
$\mathcal{R}$ to 6400\,eV and 15\% (as expected for $z \sim 8$). 
The best-fit $E_{\rm K\beta}$ value of $7090 \pm 11$~eV is consistent with 
that for Fe$^{8+}$. To summarize, we are able to explain all the Fe K-shell 
emission self-consistently with this simple two-component model using different 
ionization states and the expected K$\beta$/K$\alpha$ emissivity ratios.

\subsection{Spatial Analysis}

\begin{figure}[t]
  \begin{center}
		\vspace{2mm}
	\includegraphics[width=6cm]{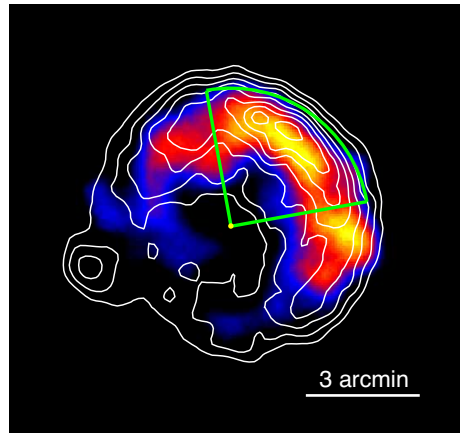}
		\vspace{2mm}
\caption{XIS image of Tycho's SNR in the 7.0--7.2\,keV (Fe-K$\beta$) band, 
where the Fe-K$\alpha$ image (same as Fig.\,\ref{fig:image1}a) is overplotted in contours.  
The synchrotron continuum flux estimated 
using Fig.\,\ref{fig:image1}b was subtracted from the raw Fe-K$\beta$ image. 
The morphology is subject to some uncertainties since a spatially-uniform 
photon index is assumed in this subtraction procedure. It is nevertheless clear 
that the Fe-K$\beta$ emission peaks at a smaller radius than the K$\alpha$ 
emission in the bright NW rim. The green sector indicates where the spatially-resolved 
spectral analysis (Fig.\,\ref{fig:spatial}) is performed. 
  \label{fig:image2}}
  \end{center}
\end{figure}

The interpretation in the previous subsection predicts that the Fe-K$\beta$ emission peaks 
interior to the K$\alpha$ emission, because the innermost ejecta were heated by the RS 
more recently and so should have a lower ionization age than the outer ejecta. 
We perform a spatial analysis to verify that this is indeed the case. 
Since the Fe-K$\beta$ emission is not as strong as the synchrotron continuum flux 
in the same energy band, subtraction of the continuum component is necessary. 
We estimate the continuum level in the 7.0--7.2\,keV (Fe-K$\beta$) band by 
scaling the synchrotron X-ray image (Fig.\,\ref{fig:image1}b) using a photon index of 2.9 
(the best-fit value for the NW rim spectrum). 
The color image in Fig.\,\ref{fig:image2} is created by subtracting this scaled data from 
the raw 7.0--7.2-keV image, where contours of the K$\alpha$ emission 
(corresponding to the image in Fig.\,\ref{fig:image1}a) are overlaid. 
As we expected, the K$\beta$ emission has a smaller peak radius than the other. 
It should be noted, however, that this imaging analysis has some uncertainties,  
because we assume that the photon index of the continuum 
X-rays is spatially uniform over the entire region. 
This is actually not the case, but the synchrotron emission in Tycho's SNR is 
known to have spatially inhomogeneous hardness \citep{Cassam07,Eriksen11}.

\begin{figure}[t]
  \begin{center}
		\vspace{2mm}
	\includegraphics[width=8.2cm]{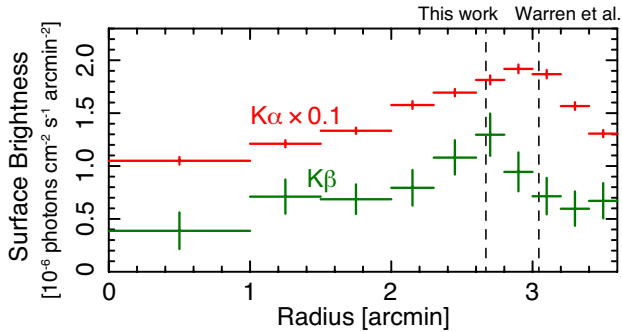}
		\vspace{1mm}
\caption{Radial profiles for the surface brightness of the Fe-K$\alpha$ and K$\beta$ 
emission in the NW quadrant, derived from the spatially-resolved spectral analysis. 
The smaller peak radius of Fe-K$\beta$ is confirmed. 
The RS positions determined by \cite{Warren05} and this work are indicated with 
the dashed lines.  
  \label{fig:spatial}}
  \end{center}
\end{figure}

This motivates our spatially resolved spectral analysis, which gives a more quantitative 
measurement of the Fe-K$\beta$ distribution. 
The azimuthal sector shown in Fig.\,\ref{fig:image2} is divided into nine radial zones of 
variable width (to account for the variation in brightness) for spectral extraction, 
by assuming the same center position for the SNR as that determined by 
the previous {\it Chandra} observations \citep{Warren05}.
The spectrum from each region is fitted with the same model applied to 
the NW rim (five Gaussians plus a power-law) allowing the photon index 
as well as the other parameters to vary freely. 
The resulting surface brightness profiles of the Fe-K$\alpha$ and K$\beta$ emission are 
shown in Fig.\,\ref{fig:spatial}, confirming the smaller peak radius of the K$\beta$ emission, 
while the K$\alpha$ peak radius is consistent with that from the {\it Chandra} data.

Our analysis demonstrates that the radial profile of the Fe-K$\beta$ emission is more 
sensitive to the immediate postshock ejecta than the K$\alpha$ emission, revealing 
a $\sim$10\% smaller RS radius than was previously determined using the Fe-K$\alpha$ 
morphology from the {\it Chandra} observations \citep{Warren05}. 
Assuming a simple shell geometry, which predicts that the maximum surface 
brightness will coincide with the inner edge of the shell \citep{Warren05}, 
we estimate the RS radius to be $7.1 \times 10^{18}\,(D/3.0\,{\rm [kpc]})$\,cm, 
or $2.3 \,(D/3.0\,{\rm [kpc]})$\,pc, where $D$ is the distance to Tycho's SNR. 
This is about 63\% of the SNR blast wave radius, $3.6 \,(D/3.0\,{\rm [kpc]})$\,pc.

\section{Interpretation}

\subsection{Comparison with Hydrodynamical Calculations}

Our new X-ray measurements have revealed that Fe in the innermost ejecta is 
in an extremely low ionization state ($z \lesssim 8$). Yet the strong Fe-K$\beta$ 
emission requires the electron temperature near the RS front be high 
enough to ionize the inner K-shell electrons of these low-$z$ ions. 
Keeping this in mind, we constrain the efficiency of collisionless electron heating 
by comparing our results with 1-D hydrodynamical simulations that incorporate 
an NEI calculation \citep{Badenes06}. For the initial conditions, we assume the ejecta 
structure expected for a typical delayed-detonation Type Ia supernova with an Fe yield 
of $\sim$$0.8M_{\odot}$ and an explosion energy of $\sim1.2 \times 10^{51}$\,ergs.  
This model reproduces the fundamental properties of Tycho's supernova: 
the historical light curve \citep{Ruiz04} and modern light echo spectrum \citep{Krause08}. 
The SNR evolution is followed to the age of Tycho's SNR assuming a uniform ambient 
density ($\rho_{\rm AM}$) of $2 \times 10^{-24}$\,g\,cm$^{-3}$. 
The result is also in good agreement with the basic dynamics (e.g., angular size, 
shock speed) of the SNR for a reasonable value of the distance \citep{Badenes06}. 
The effect of collisionless electron heating is introduced to our NEI calculations by 
parameterizing the electron-to-ion temperature ratio just behind the RS, 
$\beta = T_e/T_{ion}$.

\begin{figure}[t]
  \begin{center}
	\includegraphics[width=7cm]{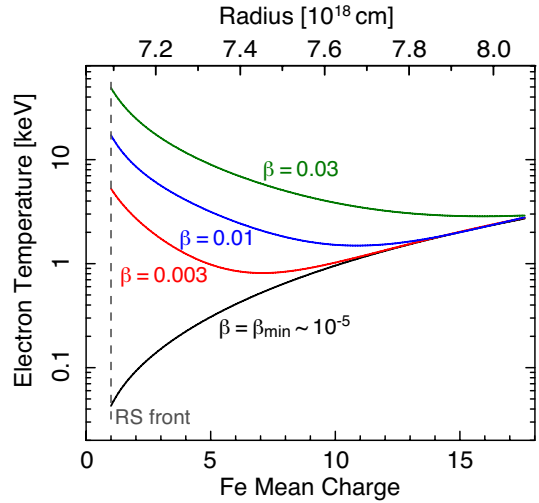}
		\vspace{2mm}
\caption{Electron temperature as a function of the mean 
charge of Fe ions from our hydrodynamical simulations. The corresponding radius 
is also given above. The black curve is the $\beta_{\rm min}$ model where no collisionless 
electron heating is assumed. The temperature ratio between the electrons and ions 
at the RS front is, therefore, set by their mass ratio. 
The models represented by the red, blue, and green curves assume that collisionless 
electron heating occurs at the RS, parametrized by ($\beta = T_e / T_{ion}$) with 
values set to 0.003, 0.01, and 0.03, respectively. 
  \label{fig:hydro}}
  \end{center}
\end{figure}

\begin{figure}[t]
  \begin{center}
	\includegraphics[width=8.2cm]{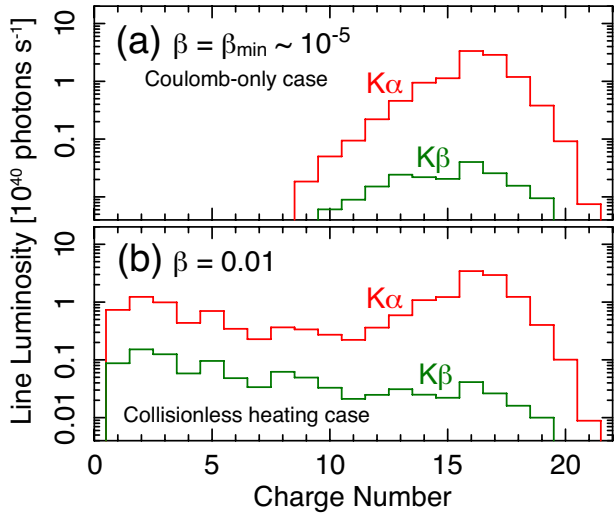}
		\vspace{1mm}
\caption{(a) Predicted luminosities of the Fe-K$\alpha$ (red) and K$\beta$ (green) 
emission lines for the $\beta_{\rm min}$ model. Owing to the low electron temperature 
in the innermost region, little or no emission from low-ionized ($z < 10$) Fe is expected. 
(b) Same as panel (a), but for the model with $\beta = 0.01$.  
Emission from Fe with various charge states, including $z < 10$ 
where higher $\mathcal{R}$ values are achieved, is expected.  
  \label{fig:luminosity}}
  \end{center}
\end{figure}

Fig.\,\ref{fig:hydro} shows the relationship between charge state, radius, and electron temperature 
for several values of $\beta$. The black curve is derived under the assumption that 
the initial temperature of each species follows the equation 
$T_i = 3\,m_i\, v_s^2/16\,k_{\rm B}$, 
which leads to $T_e/T_{\rm Fe} = m_e/m_{\rm Fe} \sim 10^{-5}$ at the RS front 
(hereafter, the $\beta_{\rm min}$ model). 
The other models, shown as the red, blue, and green curves, have larger values 
of $\beta$, which result in higher electron temperatures in the postshock region 
(hereafter, the collisionless heating models). Subsequent temperature changes 
are due to collisional processes, heating via ion--electron Coulomb collisions 
(dominant in the $\beta_{\rm min}$ model) and cooling via the ongoing collisional ionization 
process (prominent in the collisionless heating models). The $\beta_{\rm min}$ model 
predicts an electron temperature of $\lesssim$\,1\,keV in the region dominated by 
$z \leq 10$ Fe ions. The free electrons in this region are, therefore, not energetic 
enough to produce significant K-shell ionization and subsequent fluorescence. 
This is shown more quantitatively in Fig.\,\ref{fig:luminosity}a, where we plot 
the expected Fe-K$\alpha$ 
and K$\beta$ luminosity for each charge state. The $\beta_{\rm min}$ model clearly 
fails to reproduce the strong K$\beta$ emission from low-ionized Fe, in direct 
conflict with our observations.  By contrast, the collisionless heating models can 
reproduce the emission from the broad Fe ion population, as illustrated in 
Fig.\,\ref{fig:luminosity}b (corresponding to $\beta = 0.01$).

\begin{figure}[t]
  \begin{center}
		\vspace{2mm}
	\includegraphics[width=7cm]{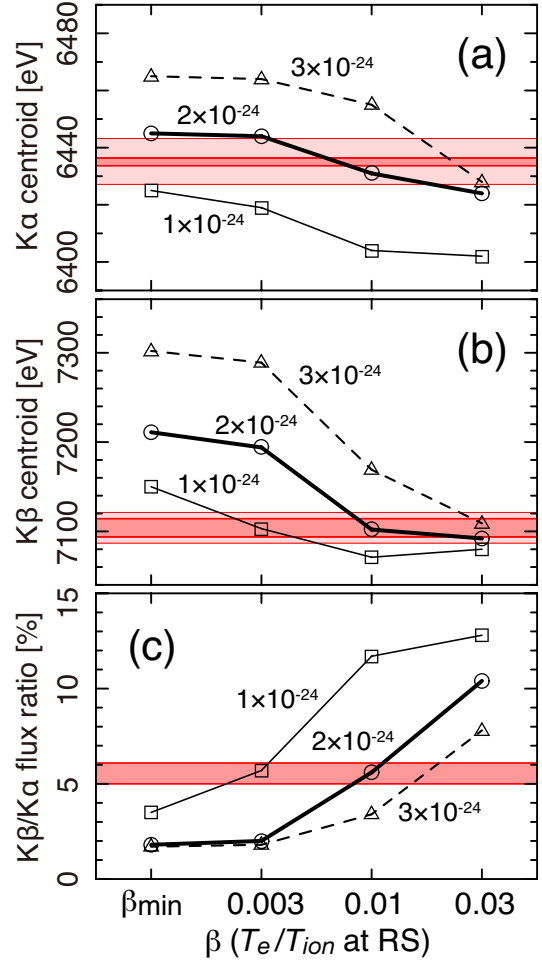}
		\vspace{1mm}
\caption{Comparison between the predicted values for (a)\,$E_{\rm K\alpha}$, 
(b)\,$E_{\rm K\beta}$, and (c)\,$\mathcal{R}$ and the observed values, for which statistical 
and systematic uncertainties are indicated with dark and light red regions, respectively. 
The thin-solid (with squares), thick-solid (with circles), and dashed (with triangles) lines 
correspond to ambient densities ($\rho_{\rm AM}$) of $1 \times 10^{-24}$, 
$2 \times 10^{-24}$, and $3 \times 10^{-24}$\,g\,cm$^{-3}$, respectively. 
We find that the model with $\rho_{\rm AM} = 2 \times 10^{-24}$ \,g\,cm$^{-3}$ 
and $\beta = 0.01$ provides the best match to the observations.
  \label{fig:comparison}}
  \end{center}
\end{figure}

In Fig.\,\ref{fig:comparison} we compare the model-predicted values of $E_{\rm K\alpha}$, 
$E_{\rm K\beta}$, and $\mathcal{R}$ to the observed values for Tycho's SNR, 
where the impact of different ambient density values, $\rho_{\rm AM}$ = 
$1 \times10^{-24}$ and $3 \times 10^{-24}$\,g\,cm$^{-3}$, are also explored. 
We confirm that the $\beta_{\rm min}$ models cannot reproduce the low 
$E_{\rm K\beta}$ and high $\mathcal{R}$ values we observe for any value of 
the ambient density. Although a lower $\rho_{\rm AM}$ value allows for a larger 
population of low-ionized Fe, it also leads to a Fe-K$\alpha$ centroid energy 
far lower than the observed value. Only the collisionless heating model with 
$\beta = 0.01$ and $\rho_{\rm AM} = 2 \times 10^{-24}$\,g\,cm$^{-3}$ explains 
all the observed values within the given uncertainties, indicating that when 
the electrons pass through the RS front, they gain an internal energy about 
$10^3$ times higher than expected without collisionless heating.

It should be noted that the recent infrared observation with {\it Spitzer} determined 
the mean ISM density around Tycho's SNR to be (2--4)\,$\times10^{-25}$\,g\,cm$^{-3}$ 
\citep{Williams13}. This is consistent with the previous estimates from the blast wave 
expansion rate \citep{Katsuda10} and the upper-limit of thermal X-ray emission 
from the shocked ISM \citep{Cassam07}, but is significantly lower than 
the value from our diagnostics. This discrepancy implies the presence of 
a density gradient around the progenitor; the SNR blast wave had initially interacted with 
relatively high density matter, which enhanced the ionization age of the outer ejecta, 
but is now expanding into the low density ISM. 
A similar interpretation was given by \cite{Dwarkadas98} and \cite{Chiotellis13}. 
Although the effect of such nonuniform ISM should be involved in our future calculations, 
we believe that this does not affect our conclusion significantly. 
The Fe-K emission from the broad range of ionization ages can be achieved 
only when the collisionless heating efficiency is high enough, 
as demonstrated in Figs.\,\ref{fig:luminosity} and \ref{fig:comparison}.

\subsection{Origin of Collisionless Electron Heating}

We have presented the first clear evidence for efficient collisionless heating of 
electrons at the RS of Tycho's SNR, which is propagating into metal-rich (no-hydrogen) 
ejecta at low magnetic field strength, with a velocity of $\sim$4000\,km\,s$^{-1}$ 
\citep{Badenes06,Hayato10}. 
As the unshocked ejecta in young Type Ia SNRs has a low temperature 
\citep[$\sim$5000\,K:][]{Hamilton88}, the Mach number is estimated 
to be a few thousands. 
Previously, the nature of collisionless heating at SNR blast waves has been studied 
via optical spectroscopy of Balmer-dominated shocks 
\citep[e.g.,][]{Raymond83,Laming96,Ghavamian01,Ghavamian07,Rakowski03,Helder11}. 
However, temperature measurements using optical spectra have been subject 
to large uncertainties in atomic cross sections, especially for high-velocity shocks 
($v_s \gtrsim 1000$\,km\,s$^{-1}$) \citep[e.g.,][]{Heng10}.

For Balmer-dominated shocks, 
two main scenarios are suggested as the origin of the electron heating: 
(a) lower hybrid wave heating in a cosmic-ray precursor \citep[e.g.,][]{Laming00,Ghavamian07}, 
and (b) plasma wave heating due to Buneman instability formed by reflected 
non-Maxwellian ions \citep[e.g.,][]{Cargill88,Matsukiyo10}. 
Both scenarios require a quasi-perpendicular shock, which is unlikely for RSs, 
where the magnetic field is expected to be quasi-parallel to the fluid flow due to 
expansion-induced stretch of the field lines. Moreover, there is little evidence for 
relativistic cosmic-rays at the RS of Tycho's SNR \citep{Warren05}, which makes 
scenario (a) unlikely in this context. 
An alternative scenario, the cross-shock potential \citep[e.g.,][]{Balikhin93}, which has also 
been suggested as the origin of energetic electrons in GRB afterglows 
\citep{Gedalin08,Sironi11}, may apply to our case. 
In this model, charge separation is created at the shock front due to the different 
gyroradii of ions and electrons, creating a potential gap where electrons arriving 
later can be accelerated. This requires no specific orientation of the background 
magnetic field, and predicts self-generation of small-scale electromagnetic fields 
\citep{Gedalin08}. 
The high Mach number for the RS also supports this analogy with GRB shocks. 
Our observation of efficient collisionless heating in the unique 
environment of an SNR RS suggests that these shocks may be fundamentally 
different from the more widely studied Balmer-dominated shocks into ISM material.


\acknowledgments

We are thankful to Drs.\ John D.\ Raymond and Timothy R.\ Kallman 
for useful information and discussion. 
This work is supported by funding from 
NASA Suzaku GO grant NNX08AZ86G (JPH)
and NASA ADP grant NNX12AF44G (RKS).

\bigskip


\bibliography{ms}

\begin{thebibliography}{}
\expandafter\ifx\csname natexlab\endcsname\relax\def\natexlab#1{#1}\fi

\bibitem[{{Badenes} {et~al.}(2005){Badenes}, {Borkowski}, \&
  {Bravo}}]{Badenes05}
{Badenes}, C., {Borkowski}, K.~J., \& {Bravo}, E. 2005, \apj, 624, 198

\bibitem[{{Badenes} {et~al.}(2006){Badenes}, {Borkowski}, {Hughes}, {Hwang}, \&
  {Bravo}}]{Badenes06}
{Badenes}, C., {Borkowski}, K.~J., {Hughes}, J.~P., {Hwang}, U., \& {Bravo}, E.
  2006, \apj, 645, 1373

\bibitem[{{Balikhin} {et~al.}(1993){Balikhin}, {Gedalin}, \&
  {Petrukovich}}]{Balikhin93}
{Balikhin}, M., {Gedalin}, M., \& {Petrukovich}, A. 1993, Physical Review
  Letters, 70, 1259

\bibitem[{{Cargill} \& {Papadopoulos}(1988)}]{Cargill88}
{Cargill}, P.~J., \& {Papadopoulos}, K. 1988, \apj, 329, L29

\bibitem[{{Cassam-Chena{\"i}} {et~al.}(2007){Cassam-Chena{\"i}}, {Hughes},
  {Ballet}, \& {Decourchelle}}]{Cassam07}
{Cassam-Chena{\"i}}, G., {Hughes}, J.~P., {Ballet}, J., \& {Decourchelle}, A.
  2007, \apj, 665, 315

\bibitem[{{Chiotellis} {et~al.}(2013){Chiotellis}, {Kosenko}, {Schure}, {Vink},
  \& {Kaastra}}]{Chiotellis13}
{Chiotellis}, A., {Kosenko}, D., {Schure}, K.~M., {Vink}, J., \& {Kaastra},
  J.~S. 2013, \mnras, 435, 1659

\bibitem[{{Dwarkadas} \& {Chevalier}(1998)}]{Dwarkadas98}
{Dwarkadas}, V.~V., \& {Chevalier}, R.~A. 1998, \apj, 497, 807

\bibitem[{{Eriksen} {et~al.}(2011){Eriksen}, {Hughes}, {Badenes}, {Fesen},
  {Ghavamian}, {Moffett}, {Plucinksy}, {Rakowski}, {Reynoso}, \&
  {Slane}}]{Eriksen11}
{Eriksen}, K.~A., {Hughes}, J.~P., {Badenes}, C., {et~al.} 2011, \apj, 728, L28

\bibitem[{{France} {et~al.}(2011){France}, {McCray}, {Penton}, {Kirshner},
  {Challis}, {Laming}, {Bouchet}, {Chevalier}, {Garnavich}, {Fransson}, {Heng},
  {Larsson}, {Lawrence}, {Lundqvist}, {Panagia}, {Pun}, {Smith}, {Sollerman},
  {Sonneborn}, {Sugerman}, \& {Wheeler}}]{France11}
{France}, K., {McCray}, R., {Penton}, S.~V., {et~al.} 2011, \apj, 743, 186

\bibitem[{{Gedalin} {et~al.}(2008){Gedalin}, {Balikhin}, \&
  {Eichler}}]{Gedalin08}
{Gedalin}, M., {Balikhin}, M.~A., \& {Eichler}, D. 2008, Physical Review E, 77,
  026403

\bibitem[{{Ghavamian} {et~al.}(2007){Ghavamian}, {Laming}, \&
  {Rakowski}}]{Ghavamian07}
{Ghavamian}, P., {Laming}, J.~M., \& {Rakowski}, C.~E. 2007, \apj, 654, L69

\bibitem[{{Ghavamian} {et~al.}(2001){Ghavamian}, {Raymond}, {Smith}, \&
  {Hartigan}}]{Ghavamian01}
{Ghavamian}, P., {Raymond}, J., {Smith}, R.~C., \& {Hartigan}, P. 2001, \apj,
  547, 995

\bibitem[{{Gu}(2008)}]{Gu08}
{Gu}, M.~F. 2008, Canadian Journal of Physics, 86, 675

\bibitem[{{Hamilton} \& {Fesen}(1988)}]{Hamilton88}
{Hamilton}, A.~J.~S., \& {Fesen}, R.~A. 1988, \apj, 327, 178

\bibitem[{{Hamilton} {et~al.}(1997){Hamilton}, {Fesen}, {Wu}, {Crenshaw}, \&
  {Sarazin}}]{Hamilton97}
{Hamilton}, A.~J.~S., {Fesen}, R.~A., {Wu}, C.-C., {Crenshaw}, D.~M., \&
  {Sarazin}, C.~L. 1997, \apj, 481, 838

\bibitem[{{Hayato} {et~al.}(2010){Hayato}, {Yamaguchi}, {Tamagawa}, {Katsuda},
  {Hwang}, {Hughes}, {Ozawa}, {Bamba}, {Kinugasa}, {Terada}, {Furuzawa},
  {Kunieda}, \& {Makishima}}]{Hayato10}
{Hayato}, A., {Yamaguchi}, H., {Tamagawa}, T., {et~al.} 2010, \apj, 725, 894

\bibitem[{{Helder} {et~al.}(2011){Helder}, {Vink}, \& {Bassa}}]{Helder11}
{Helder}, E.~A., {Vink}, J., \& {Bassa}, C.~G. 2011, \apj, 737, 85

\bibitem[{{Heng}(2010)}]{Heng10}
{Heng}, K. 2010, Publications of the Astronomical Society of Australia, 27, 23

\bibitem[{{Hwang} {et~al.}(2002){Hwang}, {Decourchelle}, {Holt}, \&
  {Petre}}]{Hwang02}
{Hwang}, U., {Decourchelle}, A., {Holt}, S.~S., \& {Petre}, R. 2002, \apj, 581,
  1101

\bibitem[{{Hwang} {et~al.}(1998){Hwang}, {Hughes}, \& {Petre}}]{Hwang98}
{Hwang}, U., {Hughes}, J.~P., \& {Petre}, R. 1998, \apj, 497, 833

\bibitem[{{Katsuda} {et~al.}(2010){Katsuda}, {Petre}, {Hughes}, {Hwang},
  {Yamaguchi}, {Hayato}, {Mori}, \& {Tsunemi}}]{Katsuda10}
{Katsuda}, S., {Petre}, R., {Hughes}, J.~P., {et~al.} 2010, \apj, 709, 1387

\bibitem[{{Krause} {et~al.}(2008){Krause}, {Tanaka}, {Usuda}, {Hattori},
  {Goto}, {Birkmann}, \& {Nomoto}}]{Krause08}
{Krause}, O., {Tanaka}, M., {Usuda}, T., {et~al.} 2008, \nat, 456, 617

\bibitem[{{Laming}(2000)}]{Laming00}
{Laming}, J.~M. 2000, \apjs, 127, 409

\bibitem[{{Laming} {et~al.}(1996){Laming}, {Raymond}, {McLaughlin}, \&
  {Blair}}]{Laming96}
{Laming}, J.~M., {Raymond}, J.~C., {McLaughlin}, B.~M., \& {Blair}, W.~P. 1996,
  \apj, 472, 267

\bibitem[{{Markevitch} {et~al.}(2005){Markevitch}, {Govoni}, {Brunetti}, \&
  {Jerius}}]{Markevitch05}
{Markevitch}, M., {Govoni}, F., {Brunetti}, G., \& {Jerius}, D. 2005, \apj,
  627, 733

\bibitem[{{Markevitch} \& {Vikhlinin}(2007)}]{Markevitch07}
{Markevitch}, M., \& {Vikhlinin}, A. 2007, Physics Reports, 443, 1

\bibitem[{{Masai}(1984)}]{Masai84}
{Masai}, K. 1984, \apss, 98, 367

\bibitem[{{Matsukiyo}(2010)}]{Matsukiyo10}
{Matsukiyo}, S. 2010, Physics of Plasmas, 17, 042901

\bibitem[{{McKee}(1974)}]{McKee74}
{McKee}, C.~F. 1974, \apj, 188, 335

\bibitem[{{Mendoza} {et~al.}(2004){Mendoza}, {Kallman}, {Bautista}, \&
  {Palmeri}}]{Mendoza04}
{Mendoza}, C., {Kallman}, T.~R., {Bautista}, M.~A., \& {Palmeri}, P. 2004,
  Astronomy \& Astrophysics, 414, 377

\bibitem[{{Meszaros} \& {Rees}(1997)}]{Meszaros97}
{Meszaros}, P., \& {Rees}, M.~J. 1997, \apj, 476, 232

\bibitem[{{Ozawa} {et~al.}(2009){Ozawa}, {Uchiyama}, {Matsumoto}, {Nakajima},
  {Koyama}, {Tsuru}, {Uchino}, {Uchida}, {Hayashida}, {Tsunemi}, {Mori},
  {Bamba}, {Ozaki}, {Dotani}, {Kohmura}, {Ishisaki}, {Murakami}, {Kato},
  {Kitazono}, {Kimura}, {Ogawa}, {Kawai}, {Mori}, {Prigozhin}, {Kissel},
  {Miller}, {Lamarr}, \& {Bautz}}]{Ozawa09}
{Ozawa}, M., {Uchiyama}, H., {Matsumoto}, H., {et~al.} 2009, \pasj, 61, 1

\bibitem[{{Palmeri} {et~al.}(2003){Palmeri}, {Mendoza}, {Kallman}, {Bautista},
  \& {Mel{\'e}ndez}}]{Palmeri03}
{Palmeri}, P., {Mendoza}, C., {Kallman}, T.~R., {Bautista}, M.~A., \&
  {Mel{\'e}ndez}, M. 2003, \aap, 410, 359

\bibitem[{{Rakowski} {et~al.}(2003){Rakowski}, {Ghavamian}, \&
  {Hughes}}]{Rakowski03}
{Rakowski}, C.~E., {Ghavamian}, P., \& {Hughes}, J.~P. 2003, \apj, 590, 846

\bibitem[{{Raymond} {et~al.}(1983){Raymond}, {Blair}, {Fesen}, \&
  {Gull}}]{Raymond83}
{Raymond}, J.~C., {Blair}, W.~P., {Fesen}, R.~A., \& {Gull}, T.~R. 1983, \apj,
  275, 636

\bibitem[{{Ruiz-Lapuente}(2004)}]{Ruiz04}
{Ruiz-Lapuente}, P. 2004, \apj, 612, 357

\bibitem[{{Sari} {et~al.}(1998){Sari}, {Piran}, \& {Narayan}}]{Sari98}
{Sari}, R., {Piran}, T., \& {Narayan}, R. 1998, \apj, 497, L17

\bibitem[{{Schwartz} {et~al.}(1988){Schwartz}, {Thomsen}, {Bame}, \&
  {Stansberry}}]{Schwartz88}
{Schwartz}, S.~J., {Thomsen}, M.~F., {Bame}, S.~J., \& {Stansberry}, J. 1988,
  Journal of Geophysical Research, 93, 12923

\bibitem[{{Sironi} \& {Spitkovsky}(2011)}]{Sironi11}
{Sironi}, L., \& {Spitkovsky}, A. 2011, \apj, 726, 75

\bibitem[{{Spitzer}(1962)}]{Spitzer62}
{Spitzer}, L. 1962, {Physics of Fully Ionized Gases}

\bibitem[{{Suh} \& {Mathews}(2000)}]{Suh00}
{Suh}, I.-S., \& {Mathews}, G.~J. 2000, \apj, 530, 949

\bibitem[{{Tidman} \& {Krall}(1971)}]{Tidman71}
{Tidman}, D.~A., \& {Krall}, N.~A. 1971, {Shock waves in collisionless plasmas}

\bibitem[{{Warren} {et~al.}(2005){Warren}, {Hughes}, {Badenes}, {Ghavamian},
  {McKee}, {Moffett}, {Plucinsky}, {Rakowski}, {Reynoso}, \&
  {Slane}}]{Warren05}
{Warren}, J.~S., {Hughes}, J.~P., {Badenes}, C., {et~al.} 2005, \apj, 634, 376

\bibitem[{{Williams} {et~al.}(2013){Williams}, {Borkowski}, {Ghavamian},
  {Hewitt}, {Mao}, {Petre}, {Reynolds}, \& {Blondin}}]{Williams13}
{Williams}, B.~J., {Borkowski}, K.~J., {Ghavamian}, P., {et~al.} 2013, \apj,
  770, 129

\bibitem[{{Wilms} {et~al.}(2000){Wilms}, {Allen}, \& {McCray}}]{Wilms00}
{Wilms}, J., {Allen}, A., \& {McCray}, R. 2000, \apj, 542, 914

\end{thebibliography}


\end{document}